\documentclass[12pt,preprint]{emulateapj}

\shorttitle{Photometry of $\delta$ Scuti stars 7~Aql and 8~Aql }
\shortauthors{Fox Machado et al.}

\begin{document}


\title{Multi-site observations of
$\delta$ Scuti stars 7~Aql and 8~Aql (a new $\delta$ Scuti
variable): the twelfth STEPHI campaign in 2003}


\author{L. Fox Machado\altaffilmark{1}, E. Michel\altaffilmark{2},
F. P\'erez Hern\'andez \altaffilmark{5,6}, J.H.
Pe\~na\altaffilmark{3},  Z.P. Li\altaffilmark{4}, M.
Chevreton\altaffilmark{2},  J.A. Belmonte\altaffilmark{5}, M.
\'Alvarez\altaffilmark{1}, L. Parrao\altaffilmark{3}, M.-A.
Dupret\altaffilmark{2}, S. Pau\altaffilmark{2}, A.
Fernandez\altaffilmark{2},  J.P. Michel\altaffilmark{2},  R.
Michel\altaffilmark{1}, A. Pani\altaffilmark{1}}


\altaffiltext{1}{Observatorio Astron\'omico Nacional, Instituto de
Astronom\'{\i}a -- Universidad Nacional Aut\'onoma de M\'exico, Ap.
P. 877, Ensenada, BC 22860, M\'exico}
\altaffiltext{2}{Observatoire de
Paris, LESIA, UMR 8109, F-92195  Meudon, France}
\altaffiltext{3}{Instituto de Astronom\'{\i}a -- Universidad
Nacional Aut\'onoma de M\'exico, Ap. P. 70-264, M\'exico, D.F.
04510, M\'exico} \altaffiltext{4}{Beijing Observatory, Chinese
Academy of Sciences, Beijing, P.R. China}
\altaffiltext{5}{Instituto de Astrof\'{\i}sica de Canarias, E-38205
 La Laguna, Tenerife, Spain}
\altaffiltext{6}{Departamento
de Astrof\'{\i}sica, Universidad de La Laguna, Tenerife, Spain}


\begin{abstract}
 We present an analysis of the pulsation behaviour of the $\delta$~Scuti stars 7~Aql (HD
174532) and 8~Aql (HD 174589) -- a new variable star --  observed in
the framework of STEPHI XII campaign during 2003 June--July. 183
hours of high precision photometry were acquired  by using
four-channel photometers at three sites on three continents during
21 days.  The light curves and amplitude spectra were obtained
following a classical scheme of multi-channel photometry.
Observations in different filters were also obtained and analyzed.
Six and three frequencies have been unambiguously detected above a
99\% confidence level in the range $190\,\mu$Hz--$300\,\mu$Hz and
$100\,\mu$Hz--$145\,\mu$Hz in 7~Aql and 8~Aql respectively. A
comparison of observed and theoretical frequencies shows that 7 Aql
and 8 Aql may oscillate with $p$ modes of low radial orders, typical
among $\delta$ Scuti stars. In terms of radial oscillations the
range of 8 Aql goes from $n=1$ to $n=3$ while for 7 Aql the range
spans from $n=4$ to $n=7$. Non-radial oscillations have to be
present in both stars as well. The expected range of excited modes
according to a non adiabatic analysis goes from $n=1$ to $n=6$ in
both stars.
\end{abstract}



\section{Introduction}

Stellar oscillations provide a powerful tool for studying the
interiors of the stars since the mode frequencies depend on the
properties of the star and give strong constraints on stellar models
and hence evolution theories. However, the observations of stellar
pulsations require extensive data sets in order to achieve
accurate frequencies and to avoid the side-lobes in the amplitude
spectrum caused by the daily cycle. Great efforts are made to
optimize the observational coverage of seismological observations,
both from the ground via multisite coordinated campaigns (e.g.
STEPHI [\citealt{michel1}], DSN [\citealt{breger}]) and from space,
where missions like COROT (\citealt{baglin}, \citealt{michel3}),
will bring a close to a 100\% observational coverage.

In the last two decades the STEPHI network (STEllar PHotometry
International) has been engaged in a long-term program aimed at
improving our knowledge and description of the physical processes at
work in the interior of $\delta$ Scuti stars. In the framework of
STEPHI campaigns most of the $\delta$ Scuti stars within Praesepe
and Pleiades clusters have been observed (e.g. \citealt{alvarez},
 \citealt{hernandez}, \citealt{fox},
 \citealt{li}) and several theoretical interpretations  have
been realized  (e.g. \citealt{michel2}, \citealt{hernandez1},
\citealt{fox1}).

The present campaign was devoted to the field star 7~Aql (HD~174532,
SAO~142696, HIP~92501), a $\delta$~Scuti variable discovered in a
systematic search and characterization of new variables, in
preparation of the COROT mission  \citep{poretti}. The star 7~Aql
has been cataloged as an evolved $\delta$~Scuti star with a low
projected rotational velocity, of $v\sin i = 32\,$km s$^{-1}$. This
star is located in the HR diagramme ($\log g = 3.8 \pm 0.1$ at
$T_{\rm{eff}} = 7400 \pm 100\,$K) in the ambiguous transition phase
between core hydrogen burning and thick shell hydrogen burning. This
phase is sensitive to the treatment of the core overshooting
process. In addition to this, its low $v\sin i$ value makes it an
interesting target for modelling and seismic interpretation, since
it restricts the seismic analysis either to a star with an intrinsic
low rotational velocity or to a faster rotator but with a low
inclination $i$ value.  For evolved $\delta$ Scuti variables a very
dense spectrum of excited modes is predicted \citep{dziembowski}.

The star 8~Aql (HD~174589, SAO~142706, HIP~92524) was chosen as
a comparison star because it is the only bright star located close
enough to 7~Aql to permit the simultaneous monitoring with the main target
within the field of view of the photometer ($\approx 12' \times
16'$) and before this campaign it was supposed a constant star
\citep{poretti}.

Table~\ref{tab:stars} shows the main observational parameters
corresponding to the target stars as taken from the SIMBAD database
operated by CDS (Centre de Donn\'ees astronomique de Strasbourg)
including the parallax measurements of HIPPARCOS. Using
these values and a reddening of $E(b-y)= 0.004$ mag
 \citep{poretti} we estimate a magnitude $M_v=1.32
\pm 0.10$ for 7~Aql and $M_v=1.42 \pm 0.10$ for 8~Aql.
 The estimated absolute magnitudes are in good
 agreement with those expected for $\delta$ Scuti variables \citep{rodriguez}.
Moreover, according to these magnitudes and colour indexs $(B-V)$ it follows that both stars are
located inside the $\delta$ Scuti instability strip.

\begin{table*}
\caption{Observational properties of the stars observed in the
STEPHI 2003 campaign. The data was taken from SIMBAD database.}
\label{tab:stars} \centering
\begin{tabular}{lcccccccc}
\hline\hline
Star & HD & HIP & ST & $V$ & $B-V$ & $v \sin i$   & parallax & $M_V$ \\
 &  &   &    &     &     & $(\mathrm{km\, s}^{-1})$ & mas & \\
\hline 7 Aql  & 174532 & 92501 & A2 & $6.9\pm 0.1$ & $+0.285\pm
0.009$ &
  $32\pm 3$    & $7.70 \pm 0.80$ & $M_v=1.32 \pm 0.10$ \\
8 Aql  & 174589 & 92524 & F2 & $6.1\pm 0.1$ & $+0.299 \pm 0.007$ &
  $105 \pm 11$ & $11.80 \pm 0.78$ & $M_v=1.42 \pm 0.10$ \\
\hline
\end{tabular}
\end{table*}

\section{Observations} \label{sec:observations}

The observations in this campaign were carried out over the period
2003 June 17--July 7. As has been done in previous STEPHI campaigns,
we observed from three sites well distributed in longitude around
the Earth: Observatorio de San Pedro M\'artir (SPM, operated by
UNAM), Baja California, Mexico; Xing Long Station (XL, operated by
the Beijing Observatory), Heibe province, China; and Observatorio
del Teide (OT, operated by the IAC), Tenerife, Spain. Thus, we are
able to limit systematic gaps in the monitoring of the light curves
of our target stars, avoiding the formation of strong aliasing
through side lobes of the spectral window in the Fourier spectrum.

Table~\ref{tab:log} gives the log of observations. Bad weather
conditions at  XL did not allow us to get more than three nights
of data from this observatory. A total of 183 hours of useful
data were obtained during 21 nights of observations from the three
sites. The overlapping between observatories was negligible and the
efficiency of the observations was 36\% of the cycle, which is
typical for a STEPHI campaign.

\begin{table}
\caption{Log of observations. Observing time is expressed in
minutes.} \label{tab:log} \centering
\begin{tabular}{cccrrr}
\hline\hline
   Day   &    Date 2003  &   HJD    & SPM & XL  & OT    \\
         &               & 2452800+ &      &     &       \\
\hline
    1    &       Jun 17  & 08    &   38  &  -   &   -   \\
    2    &       Jun 18  & 09    &   18  &  -   &   -   \\
    3    &       Jun 19  & 10    &  386  &  -   &   -   \\
    4    &       Jun 20  & 11    &   -   &  -   &   -   \\
    5    &       Jun 21  & 12    &  366  &    - &  -   \\
    6    &       Jun 22  & 13    &  359  &  -   &   -   \\
    7    &       Jun 23  & 14    &  335  &  223 &  -   \\
    8    &       Jun 24  & 15    &  359  &  -   &  297   \\
    9    &       Jun 25  & 16    &  419  &   -  &  426  \\
   10    &       Jun 26  & 17    &  422  &   -  &  425  \\
    11   &       Jun 27  & 18    &  420  &   -  &  441   \\
    12   &       Jun 28  & 19    &  369  &   -  &  436   \\
    13   &       Jun 29  & 20    &  388  &   -  &  282  \\
    14   &       Jun 30  & 21    &  425  &  204 &  44   \\
    15   &       Jul 01  & 22    &  199  &  -   &  378  \\
    16   &       Jul 02  & 23    &  411  & -    &  433 \\
    17   &       Jul 03  & 24    &  353  &  -   &  443  \\
    18   &       Jul 04  & 25    &  -    &  339 &  384   \\
    19   &       Jul 05  & 26    &  -    &  -   &  418   \\
    20   &       Jul 06  & 27    &   -   &  -   &  444    \\
    21   &       Jul 07  & 28    &  -    &  -   &  442  \\
\hline
  \multicolumn{3}{c}{Total observing time} & SPM  &   XL &  OT   \\
  \multicolumn{3}{c}{10967 (183 hours)}    & 4908 & 766 & 5293  \\
\hline
\end{tabular}
\end{table}

Four-channel photometers were used at all sites with interferometric
blue filters ($\lambda \approx 4200\,$\AA, $\Delta \lambda \approx
190\,$\AA). The channels were used to monitor the star 7~Aql,  8~Aql
and two adjacent sky background positions.
At SPM, the fourth channel was used to monitor the star 7~Aql  in a
yellow filter ($\lambda \approx 5500\,$\AA, $\Delta \lambda \approx
400\,$\AA).

\begin{figure*}
  \centering
  \includegraphics[width=17cm]{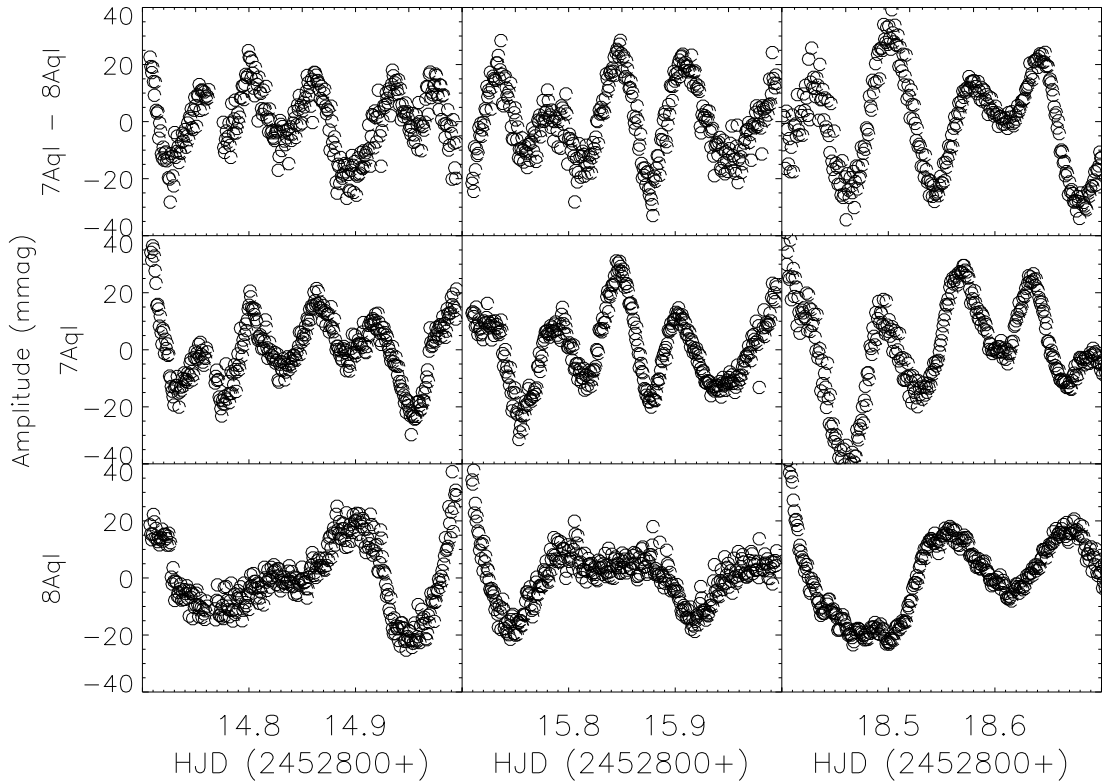}
  \caption{Examples of the light curves derived in the framework of
STEPHI XII campaign. Data averaged every 62 seconds are represented
by open circles. The three top plots correspond to the differential
light curve 7~Aql$-$8~Aql, while  the
middle and bottom plots are light curves for the star indicated in
each case. The time series were filtered by a second order polynomial.}
\label{fig:curva}
\end{figure*}

The data reduction is similar to that reported in previous STEPHI
campaigns (for details see \citealt{alvarez}). First, nightly time
series corresponding to sky background  were subtracted.
Then, we computed the magnitude differences 7~Aql$\,-\,$8~Aql and
subtracted from every light curve each night a second order
polynomial. This removed low-frequency trends that could affect the
detection of the oscillation modes at higher frequencies. Finally we
joined the data to produce one temporal series. The resulting
differential light curves for three selected days are shown in the
top panels of Fig.~\ref{fig:curva}.

Since only a comparison star was considered we also analyzed the
light curves of each star separately. The middle and bottom plots in
Fig.~\ref{fig:curva} show examples of the individual light curves of
7~Aql and 8~Aql respectively. Here a second order polynomial was
used for filtering the low frequency trends, mainly the harmonics of
the day. As can be seen in Fig.~\ref{fig:curva}, even in the case of
non-differential photometry the oscillations
in 7~Aql and 8~Aql are clearly inferred with the dominant period of 8~Aql
longer than that of 7~Aql.

With the data obtained at SPM we produced three additional temporal
series, namely, 1) 7 Aql in blue$-$8~Aql, 2): 7 Aql in
yellow$-$8~Aql, and 3) 7 Aql in blue$-$7 Aql in yellow. Thus, we are
able to search for phase shifts and amplitude ratios between the two
stars. In this case the amplitude spectrum has been filtered by a
parabola. The additional results obtained with these light curves
will be presented in Sect.~\ref{sec:colors}.

\section{Spectral analysis}\label{sec:spectral}

The frequency peaks of the light curves considered in the previous
section were obtained by performing a non linear fit to the data.
However, to show the results and for obtaining initial estimates of
the parameters, we have calculated amplitude spectra of the time
series by computing iterative sine wave fits (ISWF;
\citealt{ponman}). The figures discussed here correspond to this method.

\begin{figure}
  \resizebox{\hsize}{!}{\includegraphics{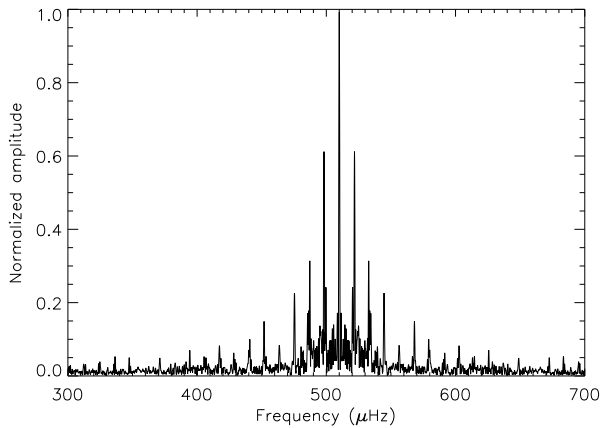}}
  \caption{Spectral window in amplitude of STEPHI XII campaign.
  First side lobes are at 58\% of the main lobe.}
  \label{fig:window}
\end{figure}

The window function of the observations is shown in
Figure~\ref{fig:window}. A one-day alias of 58\% of the main lobe
amplitude is present. The resolution as measured from the FWHM of
the main lobe in the spectral window is $\Delta \nu = 0.84 \,\mu$Hz.

\begin{figure*}
  \centering
  \includegraphics[width=17cm]{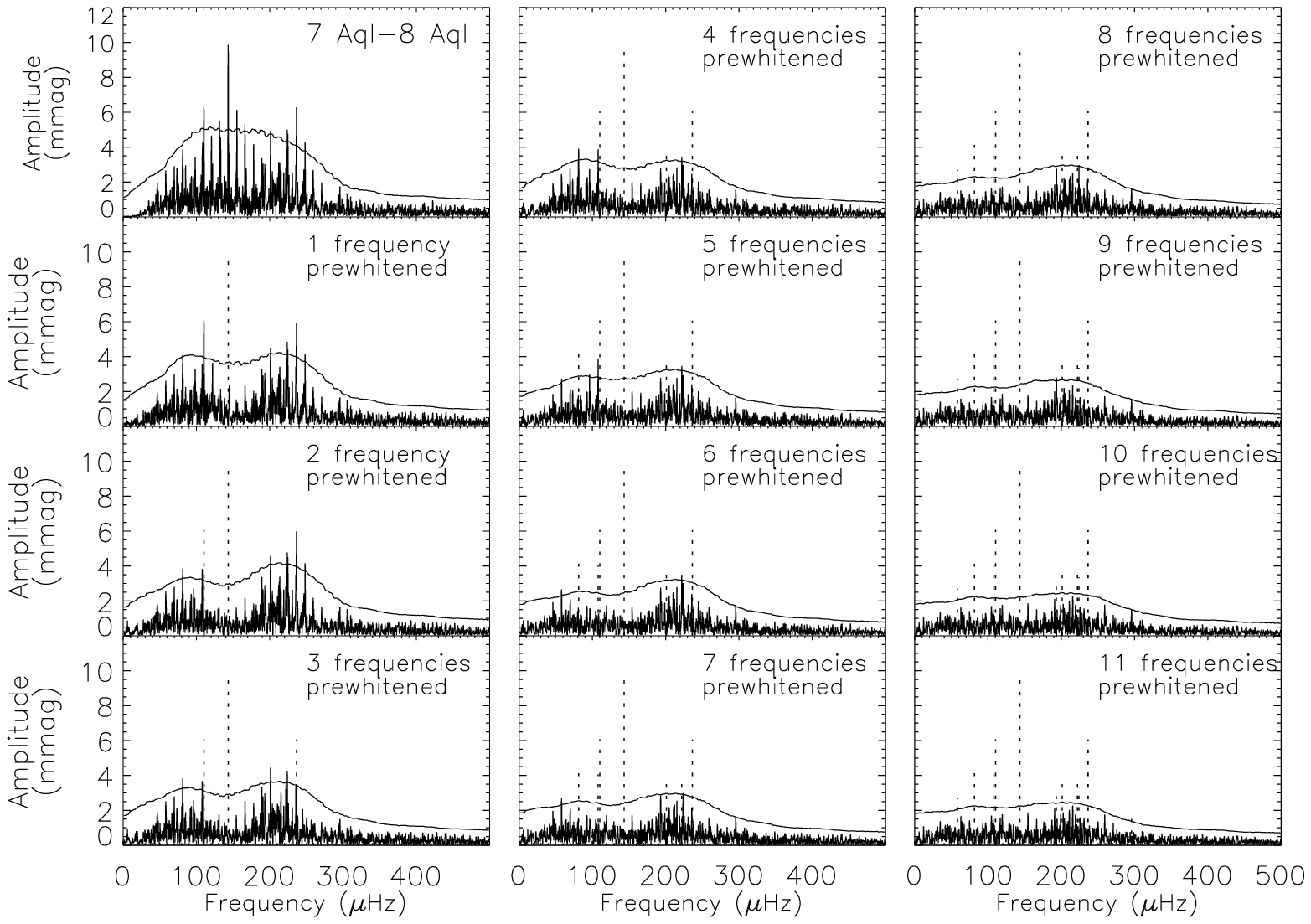}
  \caption{Pre-whitening process in the spectrum 7 Aql$-$8~Aql. In each panel,
from top to bottom, one peak above the confidence level (continuous
line) is selected and removed from the time series and a new
spectrum is obtained. In each spectrum, the prewhitened frequencies
are shown with vertical dot-dashed lines. The confidence levels are
computed as indicated in the text.}
  \label{fig:prewhite1}
\end{figure*}

The amplitude spectrum of the differential light curve 7~Aql$-$8~Aql
is plotted starting in the left top panel of Fig.~\ref{fig:prewhite1}
 and continues downward and to the right. In
order to decide which of the peaks present in the amplitude spectrum
can be regarded as signal from the star, we follow \citet{alvarez},
where it was shown that 3.7 times the mean amplitude level in the
spectrum, calculated in boxes of $100 \, \mu$Hz, can represent very
well the 99\% confidence level given by statistical tests. Similar
criteria was used in early STEPHI articles (e.g. \citealt{michel},
\citealt{fox}). This confidence level is plotted as a continuous
line in Fig.~\ref{fig:prewhite1}.

The spectrum is analyzed with a standard pre-whitening method such
that in each step the frequency peak with the largest amplitude is
subtracted from the time series. The frequency, amplitude and phase
of the pre-whitening peak is estimated simultaneously with the
previously subtracted ones by performing a non linear fit to the
original light curve. A new amplitude spectrum with all the fitted
peaks subtracted from the light curve is obtained and a new
confidence level computed. Applying the method until the whole
spectrum is below the 3.7 signal-to-noise level, the frequency peaks
which are, with a probability of 99\%, due to the star's pulsation
are obtained. The process is illustrated in
Fig.~\ref{fig:prewhite1}. It can be seen that the amplitude spectrum
of the differential light curve 7~Aql$-$8~Aql shows a spread of high
signal-to-noise peaks between $80\,\mu$Hz and $300\,\mu$Hz. The mean
noise level in the amplitude spectrum reaches $580\, \mu$mag at
$150\, \mu$Hz and $230\, \mu$mag at $400\, \mu$Hz.

\begin{figure*}
 \resizebox{\hsize}{!}{\includegraphics{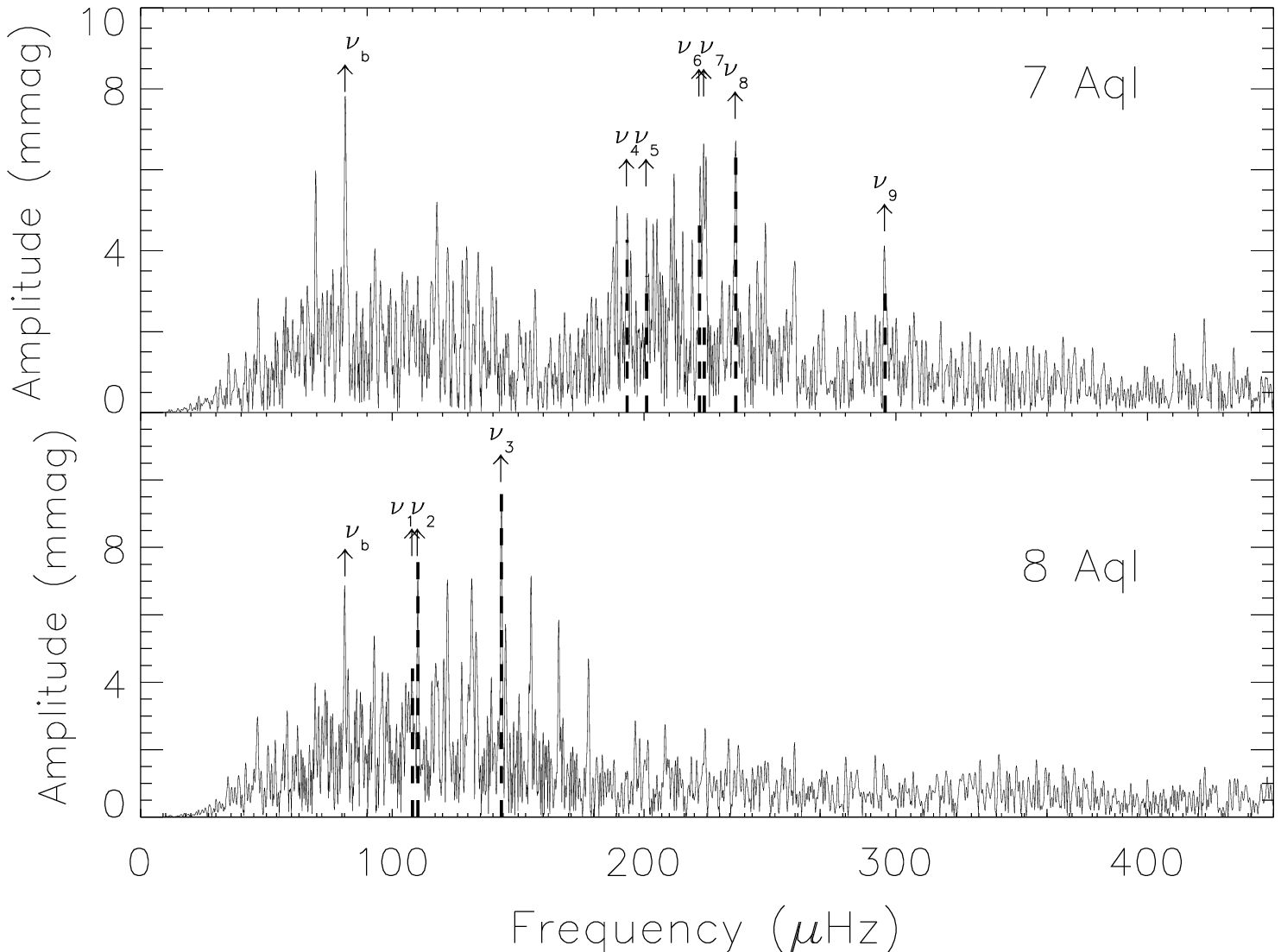}}
  \caption{Amplitude spectrum derived from individual light curves filtered by
a second order polynomial. The name of the star is indicated in each
panel.}
  \label{fig:prewhite2}
\end{figure*}

As commented earlier, both stars are variable and hence it is also
necessary to analyze the individual light curves.
In order to diminish as much as possible the transparency fluctuations on
the non-differential data we have only considered 18 nights of high photometric quality.
During those nights not only the observing conditions were good
 but also no pointing and guiding problems were present.
 A least-squares fit to a parabola was applied and
subtracted from every light curve each of the 18 nights.
The resulting amplitude spectra of the non-differential light curves
for 7~Aql and 8~Aql are
plotted in Fig.~\ref{fig:prewhite2}. The resolution in this case
is $\Delta \nu = 1.0\, \mu$Hz. The high
signal-to-noise peaks are concentrated between $190\,\mu$Hz and
$300\,\mu$Hz in 7~Aql and between $100\,\mu$Hz and $170\,\mu$Hz in
8~Aql.  Although the signal-to-noise ratio is smaller in these
spectra as compared to the differential light curve, they are good
enough to detect the oscillation frequencies due to each star as
discussed below.

\section{Results} \label{sec:results}


\subsection{Detected frequencies}

The results of the STEPHI XII multi-site campaign are summarized in
Table~\ref{tab:frequencies}, where the detected frequencies with
their corresponding amplitudes and phases are given. In total 9
peaks were detected in the spectrum of the differential light curve
---$\nu_a$ and $\nu_b$ correspond respectively to 5 and 7 c/d
and will not be considered hereafter.

\begin{table}
\caption{Frequency peaks detected above a 99\% confidence level in
the light curve 7~Aql$-$8~Aql. The origin of $\varphi$ is at
HJD~2452809.71800. $S/N$ is the signal-to-noise ratio in amplitude
after the pre-whitening process. The formal errors derived from the
(no weighting) non linear fit are indicated. Also given are the
results of the fit to the individual light curves (at fixed
frequencies). For the peaks corresponding to the light curve of 8
Aql a phase of $\pi$ have been added to help the comparison.}
\label{tab:frequencies} \centering
\begin{tabular}{lcrrrr}
\hline\hline
Series & & {\large $\nu $}  & A & $\varphi$ & $S/N$  \\
     & & ($\mu$Hz) & (mmag) &  (rad)   \\
\hline diff. & $\nu_a\mbox{}^*$ & \mbox{}58.15 &  2.7 & -- &  --  \\
             & $\nu_b\mbox{}^*$ & \mbox{}81.39 &  4.1 & --  & --  \\
& $\nu_1$ & $108.04\pm 0.05$ & $4.1\pm 0.1$ & $-3.12\pm 0.12$ &  6.9 \\
& $\nu_2$ & $110.20\pm 0.01$ & $6.1\pm 0.1$ & $-2.76\pm 0.02$ & 10.2 \\
& $\nu_3$ & $143.36\pm 0.01$ & $9.6\pm 0.1$ & $+2.72\pm 0.01$ & 15.7 \\
& $\nu_4$ & $193.28\pm 0.02$ & $2.8\pm 0.1$ & $-1.53\pm 0.04$ &  4.3 \\
& $\nu_5$ & $201.05\pm 0.01$ & $3.8\pm 0.1$ & $-2.83\pm 0.04$ &  5.6 \\
& $\nu_6$ & $222.08\pm 0.01$ & $3.6\pm 0.1$ & $-2.32\pm 0.04$ &  5.6 \\
& $\nu_7$ & $223.96\pm 0.02$ & $3.4\pm 0.1$ & $+2.30\pm 0.04$ &  5.2 \\
& $\nu_8$ & $236.44\pm 0.01$ & $6.1\pm 0.1$ & $-0.61\pm 0.02$ &  9.9 \\
& $\nu_9$ & $295.78\pm 0.03$ & $1.5\pm 0.1$ & $+1.19\pm 0.08$ &  4.1 \\
\hline 8 Aql & $\nu_1$ & 108.04 & $4.4\pm 0.2$ & $-2.91\pm 0.07$ &  2.2 \\
             & $\nu_2$ & 110.20 & $7.6\pm 0.2$ & $-3.16\pm 0.03$ &  5.1 \\
             & $\nu_3$ & 143.36 & $9.8\pm 0.2$ & $+2.81\pm 0.02$ &  7.9 \\
\hline 7 Aql & $\nu_4$ & 193.28 & $3.8\pm 0.2$ & $-1.75\pm 0.05$ &  3.6 \\
             & $\nu_5$ & 201.05 & $3.0\pm 0.2$ & $-2.83\pm 0.07$ &  2.7 \\
             & $\nu_6$ & 222.08 & $4.3\pm 0.2$ & $-1.99\pm 0.05$ &  3.9 \\
             & $\nu_7$ & 223.96 & $2.2\pm 0.2$ & $+2.42\pm 0.05$ &  2.0 \\
             & $\nu_8$ & 236.44 & $5.8\pm 0.2$ & $-0.56\pm 0.04$ &  5.5 \\
             & $\nu_9$ & 295.78 & $2.9\pm 0.2$ & $+1.14\pm 0.09$ &  3.6 \\
\hline
\multicolumn{6}{l}{\mbox{}$^*$5th and 7th harmonic of the day.}\\
\end{tabular}
\end{table}

\bigskip
In order to assign the peaks to a given star, we fit the individual
light curves to sinusoidal functions with the frequencies fixed to
the values obtained for the differential light curve. These
frequencies are marked in Fig.~\ref{fig:prewhite2} and the values of
the amplitudes and phases resulting from the fit are given in
Table~\ref{tab:frequencies}, but only for the assigned star.
Concerning this point, we note that the peaks of $\nu_2$, $\nu_3$
(in 7~Aql)  and $\nu_8$ (in 8~Aql) have $S/N>4$ as indicated in
Table~\ref{tab:frequencies} but also have a no significant amplitude
in the other star's spectrum. Concerning $\nu_4$, $\nu_6$ and $\nu_9$ they
have $S/N>3$ in the spectrum of 7~Aql and $S/N\leq 1.6$ in the
other. On the other hand, the phases are in good agreement with those of the differential light curve.

 A little more problematic can be $\nu_1$, $\nu_5$ and
$\nu_7$ since they are hidden in the noise. For $\nu_1$ and $\nu_5$
the fact that the phases agree with that of the differential data in
one of the individual light curves and also because the signal in
the other spectrum at those frequencies have $S/N \sim 1$ makes the
identification secure.

As can be seen in Fig.~\ref{fig:prewhite1} the peaks $\nu_6$ and $\nu_7$ are
close frequency pair in
the spectrum of the differential light curve.
 The existence of a
residual frequency above 99\% level, when one of them is
removed during the prewhitening process, (see Fig.~\ref{fig:prewhite1}) gives some
confidence that this pair of close oscillation frequecies are indeed intrinsic.
In fact, they have a
separation of 1.03 $\mu$Hz above the resolution limit (0.84 $\mu$Hz).
 We assigned $\nu_7$ to 7 Aql  because its
phase in the spectrum of 7 Aql is
in good agreement with that of the differential curve, in particular if one notes that
both peaks $\nu_6$ and $\nu_7$ are close enough to each other to produce a beating phenomena.

From this analysis it follows that the frequency spectrum of both
stars are not superposed. The 6 frequencies of 7~Aql are in the
frequency range $195\,\mu$Hz--$300\,\mu$Hz and have amplitudes from
$3$ to $6\,$mmag. On the other hand the 3 peaks detected in 8~Aql
are in the range $100\,\mu$Hz--$140\,\mu$Hz and with amplitudes from
$4$ to $10\,$mmag.

\subsection{Discussion of the results}

There is little information available in the literature about the
oscillation behaviour of 7~Aql and 8~Aql. In particular,
\citet{poretti} did not find photometric variability in 8 Aql. For 7
Aql they reported a ``peak-to-peak'' amplitude of 25 mmag. While this
has been enough to confirm the $\delta$ Scuti type variability in
7~Aql, the multi-periodicity of these oscillations have been
confirmed only in this campaign.

From Table~\ref{tab:frequencies} it follows that the oscillation
periods of 7~Aql and 8 Aql are within those found in others slightly
evolved $\delta$ Scuti stars observed from the ground by means of
multi-site networks (e.g  \citealt{breger}).

\subsection{Observations in different filters \label{sec:colors}}

\begin{figure}
\epsscale{.8}
\plotone{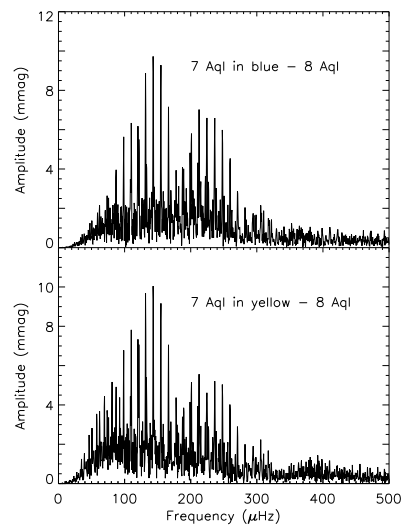}
\epsscale{1}
\caption{Amplitude spectrum derived from SPM light curves in different
filters. The name of the curves are indicated in each panel.}
  \label{fig:yellow}
\end{figure}

As was explained in Sect.~\ref{sec:observations} at San Pedro
M\'artir observatory two colour photometry was introduced in order
to test the potential help this could bring on mode identification.
Figure~\ref{fig:yellow} shows the amplitude spectrum in different
filters derived at SPM observatory.  As it is known observations
from a single site yield to a worse window function. In particular,
in SPM observatory the side-lobes in the window function are at 88\%
of the main lobe. It is also important to note that at frequencies
below $100\,\mu$Hz these spectra are dominated by the daily
aliasing, mainly for the second one where we performed differential
photometry with two stars in different filters.

In order to determine the differences in amplitude and phase between
the different spectra a linear least-squares fit was performed to
the SPM light curves with the frequencies fixed to the set listed in
Table~\ref{tab:frequencies}. The resulting phase differences and
amplitude ratios between curves 7 Aql in blue$-$8 Aql and 7 Aql in
yellow$-$8 Aql are listed in Table~\ref{tab:dif} (columns 3--4). The
given uncertainties correspond to the result of the no weighting
fit. Since these uncertainties usually underestimate the true
errors, it is not clear that the resulting differences between the
two filters are real or are an artifact consequence of the bad
coverage from a single site. To test this point we have also
computed the spectrum of the light curve 7~Aql in blue$-$7~Aql in
yellow. In this case, the extinction effects are not properly
canceled and the daily aliasing dominate the spectrum. Even so we
find some stellar signal above noise and the results of the linear
fit at fixed frequencies are given in Table~\ref{tab:dif} (columns
5--7). At least for $\nu_8$ we obtain a significant $S/N$, hence
showing that the differences in amplitude and phase between the
filters considered are indeed real.

\begin{table*}
\caption{Amplitude ratios $A_{v}/A_{y}$ and phase differences
$\varphi_v-\varphi_y$ between curves 7 Aql in blue$-$8 Aql and 7 Aql
in yellow$-$8 Aql. Also shown are the amplitudes, phases and
signal-to-noise ratios, $S/N$, obtained from a fit to the light
curve 7~Aql in blue$-$7~Aql in yellow. The quoted errors are from
the no weighting least squared fit.} \label{tab:dif} \centering
\begin{tabular}{ccccccc}
\hline\hline \multicolumn{2}{c}{Frequency} & $A_{v}/A_{y}$ &
$\varphi_v-\varphi_y$ &
  $A_{v-y}$ & $\varphi_{v-y}$ & $S/N$ \\
 & ($\mu$Hz) &               &       (radians)       &
   (mmag)   &    (radians)   &  \\
    \hline
 $\nu_{4}$  & 193.28 & $1.42 \pm 0.15$ & $-0.08 \pm 0.11$ &
 --              & --                & 1.6 \\

 $\nu_{5}$  & 201.05 & $0.99 \pm 0.08$ & $-0.18 \pm 0.08$ &
 --              & --                & 1.3 \\

 $\nu_{6}$  & 222.08 & $1.07 \pm 0.06$ & $+0.01 \pm 0.06$ &
 --              & --                & 0.5 \\

 $\nu_{7}$  & 223.96 & $1.64 \pm 0.24$ & $+0.36 \pm 0.14$ &
 $1.31 \pm 0.14$ & $+2.66 \pm 0.10$  & 2.8 \\

 $\nu_{8}$  & 236.45 & $1.25 \pm 0.05$ & $-0.01 \pm 0.04$ &
 $1.39 \pm 0.11$ & $-0.67 \pm 0.08$  & 3.2 \\

$\nu_{9}$  & 295.78 & $1.76 \pm 0.13$ & $-0.19 \pm 0.08$ &
 -- & --  & 2.1 \\

     \hline
  \end{tabular}
\end{table*}

The derived values of amplitude ratios and phase shift are discussed
in the next section.

\section{Comparison with theoretical models}

In this section  we will compute a set of representative models for
our target stars to perform simple frequency comparisons, which will
allow us to obtain some insights on their pulsation behaviour.

 The computation of the theoretical evolutionary sequences and the
calibration to the Johnson photometric system are explained in
\citet{fox1}. In particular, we used the CESAM evolution code
\citep{morel} with input physics appropriate to $\delta$ Scuti stars
and with a chemical initial composition of $Z=0.02$ and $Y=0.28$.
Models with and without convective overshooting have been considered, in the
latter case being the parameter $\alpha_{\rm ov}=0.2$. Rotating
evolutionary models were computed just for modelling 8 Aql since it is
a rapid rotator ($v\sin i = 105\,$km s$^{-1}$).

\begin{figure}
\resizebox{\hsize}{!}{\includegraphics{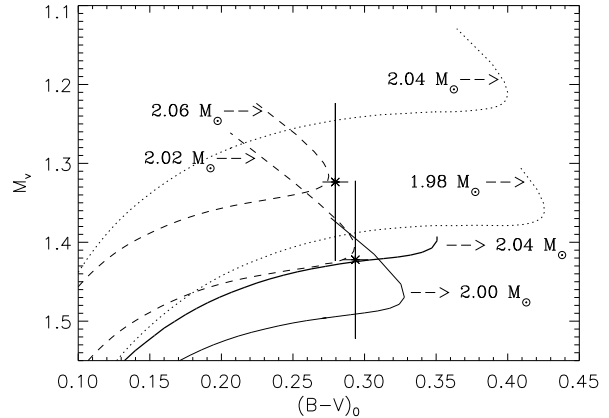}}
\caption{Colour--magnitude diagram showing the location of the
target stars. The slightly cooler star is 8 Aql. Evolutionary sequences
of non-rotating models with and without convective overshooting are shown by
dotted and dashed lines respectively. The continuous lines are
evolutionary sequence of models with no convective overshooting and initial
rotation velocities of $v \simeq 121\,$km s$^{-1}$(thin line) and $v
\simeq 180\,$km s$^{-1}$ (thick line).  The error bars give the position of
the stars according to the uncertainties listed in
Table~\ref{tab:stars}.} \label{fig:models}
\end{figure}

Figure~\ref{fig:models} shows the de-reddened position of the target
stars in a colour-magnitude diagram. The slightly brighter star is 7
Aql. The error bars were already given in Table~\ref{tab:stars} and
have been derived using HIPPARCOS data. The dashed and dotted lines
are  evolutionary sequences of non-rotating models  with and without
convective overshooting  respectively, giving a range of masses suitable for 7
Aql. The continuous lines correspond to  evolutionary tracks of
rotating models of 2.00 $M_{\odot}$ (thin line) and 2.04 $M_{\odot}$
(thick line) which match approximately the observational position of
8 Aql.  In this case, the initial rotational velocity used in each
evolutionary sequence (121 km s$^{-1}$ for 2.0 $M_{\odot}$ and 180 km s$^{-1}$ for
2.04 $M_{\odot}$) is consistent with the high rotation rate observed
in this star. Solid-body rotation, with
 conservation of global angular momentum during the evolution were assumed.

According to the models depicted in Fig.~\ref{fig:models} our target
stars could be in a similar evolutionary stage. In particular, their ages
should be between 760 and 1100 Myr with a range of masses of 2.00
$\pm$ 0.04 $M_{\odot}$.

We now use these models to derive a range of radial orders of the
target stars.  The adiabatic eigenfrequencies were computed using
the code FILOU (\citealt{tran}, \citealt{suarez}). As can be seen in
Table~\ref{tab:frequencies}, some of the frequencies of 7~Aql (e.g.
$\nu_1$ and $\nu_2$) and 8~Aql (e.g. $\nu_6$ and $\nu_7$) are so
close that non radial oscillations need to be present. This is a
common result among $\delta$~Scuti stars. However, at the
evolutionary stage of our target stars the non radial oscillation
modes develop a mixed character in the frequency range of interest
making it difficult to assign radial orders to them without
additional constraints.
 For this reason we prefer to give only the range
of $n$ associated to the radial oscillations, keeping in mind that
$p$ modes with $\ell>0$ are present as well.

In Table~\ref{tab:modes} the possible range of radial orders of
$\ell=0$ modes for the observed frequencies is given. Here we have
considered both models with and without convective overshooting.
 The theoretical frequencies of 8 Aql
were computed up to second order in the rotation rate perturbative
treatment. We note that the evolutionary sequences of rotating models
with $\alpha_{\rm ov}=0.2$ are not shown in Fig.~\ref{fig:models},
but were computed with the same masses as indicated by the
continuous lines.

\begin{table}
\caption{For each observed frequencies the possible range of radial
orders of $\ell=0$ modes is given. All the models have $Z=0.02$ and
$Y=0.28$.} \label{tab:modes} \centering
\begin{tabular}{lcc}
\hline\hline & \multicolumn{1}{c}{$\alpha_{\rm ov} = 0.00$} &
\multicolumn{1}{c}{$\alpha_{\rm ov} = 0.20$} \\
\hline
           &$\ell=0$&$\ell=0$\\
 \hline
Obs. frec. & $n$  & $n$  \\
($\mu$Hz) \\
\hline
\\
8 Aql \\
$\nu_{1}=108.04$ & 1,2  & 1,2    \\
$\nu_{2}=110.19$ & 1,2 &  1,2    \\
$\nu_{3}=143.36$ & 2,3 &  2,3  \\
\\
7 Aql \\
$\nu_{4}=193.29$ & 4 &  4,5      \\
$\nu_{5}=201.05$ & 4 &4,5      \\
$\nu_{6}=222.08$ & 5    & 5,6      \\
$\nu_{7}=223.96$ & 5   & 5,6      \\
$\nu_{8}=236.45$ & 6   & 6,7     \\
$\nu_{9}=295.78$ & 7   & 7 \\
                    &         &  \\
\hline
\end{tabular}
\end{table}

Finally we have performed preliminary non-adiabatic computations
corresponding to our target stars. The non-adiabatic pulsation
models considered include the Time-Dependent Convection (TDC)
treatment of \citep{gabriel} and \citep{grigahcene}.  The probable
evolutionary stage of the stars is shown in Fig.~\ref{fig:models}.
In particular, for 7 Aql we have considered structure models with
$M=2M_{\sun}$, $T_{\rm eff}=7400$~K, $\log(L/L_{\sun})=1.387$,
$X=0.7$, $Z=0.02$, $\alpha_{\rm ov}=0.2$ and different values of the
mixing length parameter $\alpha$ (1.8, 1.5, 1 and 0.5).

The stability analysis shows that all the modes in the observed
range of frequencies are predicted to be overstable, for 7~Aql and
8~Aql. The whole range of predicted overstable radial modes goes
from $n=1$ to $n=6$.

\begin{figure}
\resizebox{\hsize}{!}{\includegraphics{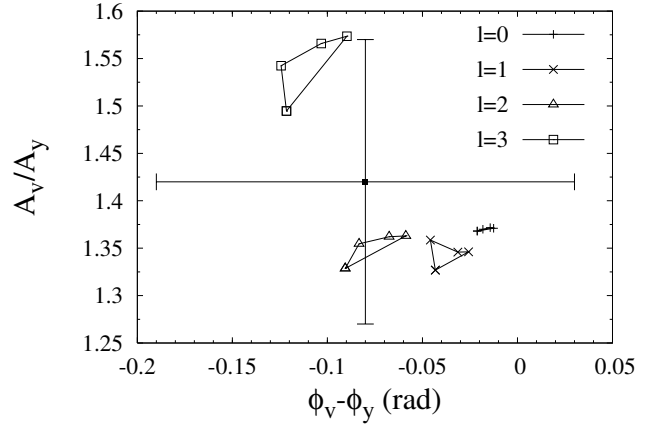}}
\caption{Theoretical amplitude ratios and phase differences compared
with observations (big cross) for the mode $\nu_4$ of 7~Aql. Each
group of points related by lines corresponds to a different degree
$\ell$. Each point of a group corresponds to a model with given
$\alpha$ (values: 0.5, 1, 1.5, 1.8).} \label{fig:difphase}
\end{figure}

As shown by \citet{dudsd05}, our TDC treatment allows a much more
secure multi-colour photometric identification of the degree $\ell$
of the modes in $\delta$~Sct stars. However, in the case of 7~Aql,
the observed error bars on the amplitudes and phases are
too large to allow this identification. This is illustrated in
Fig.~\ref{fig:difphase}, where the theoretical amplitude ratios and
phases obtained with our TDC treatment for models with different
$\ell$ and $\alpha$ are compared with observations (large cross).

\section{Conclusions}

We have presented the results obtained in the STEPHI XII multi-site
campaign. The stars 7 Aql and 8 Aql were monitored for a period of
21 days during 2003 June--July.  The analysis reveals that
8 Aql is a new $\delta$ Scuti variable.

  The three-continent run allowed us
to reach a low noise level ($\sim 230\,\mu$mag at $400\,\mu$Hz) and
a good spectral window (first side lobes at 58\% of the main lobe in
amplitude). The efficiency of the observations was 36\% of the
cycle. In fact our campaign represents the most extensive work on 7
Aql and 8 Aql in terms of the time, data points and observatories
involved.

 A long differential light curve of 7~Aql$-$8~Aql was obtained.
 High quality photometric nights of data were used to produce non-differential
time series of each star.
We have disentangled the peaks present in each star
by comparing the amplitude and phases values of the non-differential
time series to that of the differential light curve at fixed frequencies values.
 We have found that the frequency spectrum of
both stars are not superposed.
7~Aql and 8~Aql have been found to be multi-periodic pulsators with
at least six and three modes of oscillations respectively.
The resulting amplitude spectra of
non-differential and differential photometry do not shed any doubt on the results.
Even so, we are considering new observations involving
the faint comparison stars close to 7 Aql and 8 Aql
in order to confirm the list of oscillation frequencies
by means of ensemble photometry.

A comparison of observed and theoretical frequencies of the radial
modes reveals that pulsations in 7 Aql and 8 Aql can be due mostly
to low order $p$ modes with radial orders typical among $\delta$
Scuti stars. In particular, we have found that oscillation spectrum
of 7 Aql and 8 Aql contain frequencies of radial modes with
overtones up to $n=7$ and $n=3$ respectively. Non-radial
oscillations must be present in both stars as well.  A non-adiabatic
analysis shows that the modes in the observed range of frequencies
in 7~Aql and 8~Aql are theoretically overstable.
 The same range of radial orders were
expected to be excited in both stars as opposite to what is found
in the observations. In particular, the whole range of predicted overstable radial modes
goes from $n=1$ to $n=6$ in both stars, whereas
the observed frequency peaks in each star span over a more restricted range
of consecutive radial orders between $n=1$ to $n=3$ (in 8 Aql)
and between $n=4$ to $n=7$ (in 7 Aql).

\acknowledgments

This work has received financial support from the French CNRS, the
Spanish DGES (AYA2001-1571, ESP2001-4529-PE and
ESP2004-03855-C03-03), the Mexican CONACYT and UNAN under grant
PAPIIT IN110102 and IN108106, the Chinese National Natural Science
Foundation under grant number 10573023 and 10433010. Special thanks
are given to the technical staff and night assistant of the Teide,
San Pedro M\'artir and Xing-Long Observatories and the technical
service of the Meudon Observatory. The 1.5 m Carlos S\'anchez
Telescope is operated on the island of Tenerife by the Instituto de
Astrof\'{\i}sica de Canarias in the Spanish Observatorio del Teide.
This research has made use of the SIMBAD database operated at CDS,
Strasbourg (France). We thank the anonymous referee for helping us
to improve the manuscript.

\end{document}